\documentstyle[prl,preprint,aps]{revtex}
\input{epsf}
\tighten

\begin{document}

\draft

\title
{Charge transfer electrostatic model of
compositional order in perovskite alloys.}

\author
{Zhigang Wu and Henry Krakauer}

\address
{Department of Physics, College of William and Mary\\
Williamsburg, VA 23187}

\date{\today}
\maketitle

\begin{abstract}
We introduce an electrostatic model including charge transfer, which
is shown to account for the observed B-site ordering in Pb-based
perovskite alloys. 
The model allows charge transfer between A-sites
and is a generalization of Bellaiche and Vanderbilt's purely
electrostatic model. The large covalency of Pb$^{2+}$ compared to
Ba$^{2+}$ is modeled by an environment dependent effective A-site
charge. 
Monte Carlo simulations of this
model successfully reproduce the long range compositional
order of both Pb-based and Ba-based complex A(BB$^{'}$B$^{''}$)O$_3$ 
perovskite alloys. 
The models are also extended to study systems with A-site 
and B-site doping, such as 
(Na$_{1/2}$La$_{1/2}$)(Mg$_{1/3}$Nb$_{2/3}$)O$_3$,
(Ba$_{1-x}$La$_{x}$)(Mg$_{(1+x)/3}$Nb$_{(2-x)/3}$)O$_3$ and
(Pb$_{1-x}$La$_{x}$)(Mg$_{(1+x)/3}$Ta$_{(2-x)/3}$)O$_3$. 
General trends are reproduced by purely
electrostatic interactions, and charge transfer
effects indicate that local structural
relaxations can tip the balance between different B-site
orderings in Pb based materials.
\end{abstract}

\pacs{PACS numbers: 64.60.Cn, 81.30.-t, 77.84.-s, 77.84.Dy}


\section{Introduction}

Complex solid solution perovskite structure based relaxor
ferroelectrics such as Pb(Zr$_{1\-x}$Ti$_x$)O$_{3}$ (PZT) and
\mbox{$(1-x)$~Pb(Zn$_{1/3}$Nb$_{2/3}$)O$_{3}$~+~$x$~PbTiO$_{3}$
(PZN-PT)} show great promise as piezoelectric
transducers. \cite{piezo_devices} In their simplest form, PZN-PT
materials are B-site alloys of the form A(BB$^{'}$B$^{''}$)O$_3$, with
three different B-site cations from group II, IV, and V, which exhibit
compositionally dependent B-site atomic ordering. For example, at
1640~$^{\circ}$C, when the tetravalent composition $x$ is increased in
(1-$x$)~Ba(MgNb)O$_{3}$~+~$x$~BaZrO$_{3}$ (BMN-BZ) alloy, the
following sequence of B-site ordering is observed: [111]$_{1:2}$ order
for $x<5\%$; then [111]$_{1:1}$ order for $5\%<x<25\%$; and finally
disorder for larger $x$. \cite{ad} Other Ba-based perovskites, e.g.,
(1-$x$)~Ba(Mg$_{1/3}$Ta$_{2/3}$)O$_{3}$~+~$x$~BaZrO$_{3}$ (BMT-BZ),
\cite{ad} (1-$x$)~Ba(Mg$_{1/3}$Nb$_{2/3}$)O$_{3}$~+~$x$~BaZrO$_{3}$
(BMN-BZ), \cite{cadp} display a similar sequence of B-site order.  On
the other hand, for Pb-based systems, e.g.,
(1-$x$)~Pb(Mg$_{1/3}$Ta$_{2/3}$)O$_{3}$~+~$x$~PbZrO$_{3}$ (PMT-PZ),
[111]$_{1:2}$ order is not observed at $x$~=~0; instead, annealing
between 1325$^{\circ}$C and 1350$^{\circ}$C results in [111]$_{1:1}$
order all the way down to $x$~=~0. \cite{ad2,ad3} Other Pb-based
perovskites , e.g., Pb(Mg$_{1/3}$Nb$_{2/3}$)O$_3$ (PMN),
\cite{Husson1,Husson2,Chen} display similar B-site ordering.

Chemical substitutions on the A-site can also strongly afftect B-site
ordering. For example,
(Na$_{1/2}$La$_{1/2}$)(Mg$_{1/3}$Nb$_{2/3}$)O$_3$ shows [111]$_{1:1}$
B-site ordering at all measured temperatures; and 
[001]$_{1:1}$ A-site ordering occurs
below $\sim
950^{\circ}$C. \cite{dcd} In
(Ba$_{1-x}$La$_{x}$)(Mg$_{(1+x)/3}$Nb$_{(2-x)/3}$)O$_3$, [111]$_{1:1}$
B-site ordering is induced and coexists with [111]$_{1:2}$ order as
low as 5\% La doping on A-sites. For $x \geq 0.1$, [111]$_{1:1}$
B-site ordering is stablized. \cite{ad4} Similar B-site ordering
sequences are observed for
(Ba$_{1-x}$La$_{x}$)(Zn$_{(1+x)/3}$Nb$_{(2-x)/3}$)O$_3$. \cite{ad5} On
the other hand, annealing
(Pb$_{1-x}$La$_{x}$)(Mg$_{(1+x)/3}$Ta$_{(2-x)/3}$)O$_3$ at 1300
$^{\circ}$C yields [111]$_{1:1}$ B-site ordering for all $x$, and the
strongest 1:1 reflections are measured at $x = 0.5$. \cite{mad}
(Pb$_{1-x}$La$_{x}$)(Mg$_{(1+x)/3}$Nb$_{(2-x)/3}$)O$_3$ displays
similar B-site ordering sequence. \cite{ad2000}

Bellaiche and Vanderbilt (BV) introduced an electrostatic model that
was remarkably successful in explaining the B-site ordering of
A$^{+2}$(BB$^{'}$)O$_3$ perovskite alloys, using only electrostatic
interactions between B-site. \cite{bv} However, the model failed to
correctly describe the small $x$ behavior of PMT-PZ described above. The
BV model has also not been extended to study the effects of A-site
chemical substitution.

The different behavior of Ba- and Pb-based alloys may be presumed to
arise from the greater covalency of Pb compared to Ba. 
First-principles calculations for Ba based alloys show that typical
B-site stacking structural energy differences can be as large as 60
kJ/(mol-ABO$_3$). \cite{bc} This is an order of magnitude larger than
in similar Pb based alloys, \cite{bc,wensell99,wensell-2000} and reflects
the greater covalency of Pb compared to Ba.
In this paper,
we investigate modeling Pb covalency through the
inclusion of a local-configuration dependent A-site charge, {\it i.e.}
charge transfer between A-sites.  The resulting electrostatic models
with and without charge transfer are used to study B-site atomic
ordering in undoped and La-doped Ba- and Pb-based perovskite
alloys. 

\section{Electrostatic Models Without Charge Transfer}

We will be considering supercells of the ABO$_3$ perovskite structure
with possible mixtures of different atomic species on the A and B
sites. Strain and relaxation of atomic positions are neglected in all
that follows. The total electrostatic energy is then given by
\begin{equation}
 E={e^2\over2} \sum_{(l\tau)\ne(l'\tau')}
   {Q_{l\tau}Q_{l'\tau'}\over\epsilon\,
     |{\bf R}_{l\tau}-{\bf R}_{l'\tau'}|} \;,
\end{equation}
where $l$ indexes the primitive cubic cell within the supercell,
$\tau$ indicates the sublattice position in the $l^{th}$ primitive cell
($\tau$~=~\{A,B,O$_1$,O$_2$,O$_3$\}), and $\epsilon$ is the electronic
dielectric constant. The formal oxygen ionic charge $Q_{l,\rm O}=-2$
will be used in the following, and the A- and B-site charges
will be referenced to PbTiO$_3$:
\begin{equation}
 Q_{l,\rm A}= 2+\Delta q_{l,\rm A} \;,
\label{qaeq}
\end{equation}
and
\begin{equation}
 Q_{l,\rm B}= 4+\Delta q_{l,\rm B} \;,
\label{qbeq}
\end{equation}
{\it e.g.} $\Delta q_{l,\rm A}$ = 0, -1, +1 for Ba, Na, and La
respectively, and $\Delta q_{l,\rm B}$ = 0, -2, +5 for Ti, Mg, and Nb
respectively, give the formal ionic charges for these atoms.  BV
showed that up to a constant, the configurationally averaged total
energy depends only on terms quadratic in the $\Delta q_{l,\tau}$. The
configurational electrostatic energy can then be written as:
\begin{equation}
 E_{\rm AB}={e^2\over2 \epsilon} 
  \left [ 
   \sum_{(lA)\ne(l'A')} 
      {\Delta q_{lA}\Delta q_{l'A'}
      \over\, |{\bf R}_{lA}-{\bf R}_{l'A'}|} 
   + 
   \sum_{(lB)\ne(l'B')} 
      {\Delta q_{lB}\Delta q_{l'B'}
      \over\, |{\bf R}_{lB}-{\bf R}_{l'B'}|} 
   + 
   2 \sum_{lA,l'B} 
      {\Delta q_{lA}\Delta q_{l'B}
    \over\, |{\bf R}_{lA}-{\bf R}_{l'B}|} 
  \right ] \;,
\label{ABeq}
\end{equation}

BV introduced this model for
A$^{+2}$(BB$^{'}$)O$_3$ perovskite alloys. 
Since they only
considered alloys for which A-sites are occupied by Ba or Pb, where
$\Delta q_{l,\rm A}$~=~0, the configurational electrostatic energy
then depends only on the B-sublattice:
\begin{equation}
 E_{\rm B} = {e^2\over2\epsilon a} \sum_{l\ne l'}
    {\Delta q_{l,\rm B} \Delta q_{l^{'},\rm B}\over |\bf l-l^{'}|} \;,
\label{bveq}
\end{equation}
where $a$ is the cubic lattice constant, and ${\bf R}_l={\bf l}a$.

This model was
remarkably successful in explaining the B-site ordering of
A$^{+2}$(BB$^{'}$)O$_3$ perovskite alloys using Eq. (\ref{bveq}). \cite{bv} 
Using a B-site-only notation to
describe the different classes of alloys simplifies the discussion. Thus,
IV$_{x}$IV$^{'}_{1-x}$ denotes a homovalent B-site binary alloy having
tetravalent B-atoms, e.g., Pb(ZrTi)O$_{3}$; and
II$_{(1-x)/3}$IV$_{x}$V$_{2(1-x)/3}$ indicates a heterovalent B-site ternary,
such as (1-$x$)~Ba(Mg$_{1/3}$ Nb$_{2/3}$)O$_{3}$~+~$x$~BaZrO$_{3}$. 
The BV model predicted the following B-site orderings \cite{bv}: 
(1) For III$_{(1-x)/2}$IV$_x$V$_{(1-x)/2}$ and
II$_{(1-x)/2}$IV$_x$VI$_{(1-x)/2}$ heterovalent ternaries, a
rocksalt-type [111]$_{1:1}$ ordering becomes disordered with
progressively increasing $x$, consistent with experimental
observations.  
(2) For II$_{(1-x)/3}$IV$_x$V$_{2(1-x)/3}$ heterovalent
ternaries, the model predicts a richer phase diagram, with
[111]$_{1:2}$ order for $x$ less than about 5\%, [111]$_{1:1}$ order
for $x$ greater than 5\%, followed by disorder for large $x$.  This is
in good agreement with observations for Ba based systems but disagrees
for Pb based systems, where [111]$_{1:1}$ order persists down to
$x$~=~0. They proposed a possible explanation, \cite{bv} postulating a
small amount of Pb$^{4+}$ on the B-sublattice to explain the
stabilization of the 1:1 phase.  As far as we know, there is no
evidence of the presence of Pb$^{4+}$ on the B-sublattice.

\section{Charge Transfer Electrostatic Model}

As mentioned, first-principles calculations for Ba based alloys show that typical
B-site stacking structural energy differences can be as large as 60
kJ/(mol-ABO$_3$), \cite{bc} which is an order of magnitude larger than
in the Pb based alloys. \cite{bc,wensell99,wensell-2000} The energy
differences in the Ba based compounds can be approximately reproduced
in the BV model by choosing the value of its one free
parameter $\epsilon$~=~10 at the experimental volume. \cite{bv,bc} 
However, the BV model is then
unable to account for the much smaller energy differences in the Pb
based compounds. 
The greater covalency of Pb compared to Ba permits
larger local relaxations, reducing the energy differences between
different Pb based B-site stackings. Covalent interactions
are of course absent in a purely electrostatic model. 

We attempt to model the greater Pb covalency by
allowing charge transfer between A-sites, depending on the local
B-site configuration.  
Our charge transfer electrostatic model is closely patterned after
similar models that were introduced to model the configuration
dependence of Coulomb energies for pseudobinary
alloys,\cite{Schilfgaarde86,wei87,mwz} such as (Al$_{0.5}$Ga$_{0.5}$)As.
These calculations found that Coulomb contributions of calculated
mixing enthalpies were commensurate with that of lattice mismatch
energies. \cite{Schilfgaarde86} In addition, the trend of stability of
(AC)$_n$(BC)$_n$ superlattices predicted by these models was 
consistent with first-principles total-energy calculations for
lattice-matched semiconductor superlattices. \cite{mwz}

To avoid confusion in the following, we use an
overhead ``$\sim$'' to denote charges that depend on the local
configuration of a site, while charges without the overhead $\sim$
denote ionic charges that depend only on the atom occupying the site.
The simplest way to include charge transfer in the model is to define a
nearest-neighbor configuration-dependent effective A-site charge
$\widetilde{Q}_{l,\rm A}$:
\begin{equation}
 \widetilde{Q}_{l,\rm A}=q_{\rm A}+\Delta \widetilde{q}_{l,\rm
 A}=2+\Delta \widetilde{q}_{l,\rm A} \;.
\end{equation}
where
\begin{equation}
 \Delta \widetilde{q}_{l,\rm A}=-\lambda({1\over8} \sum_{i=n.n. {\rm B-sites}}
                 {\Delta q_{i,\rm B} }) \;.
\label{lambdaeq}
\end{equation}
The parameter $\lambda$ controls the charge transfer 
between  the A-sites, {\it i.e.} the degree of covalency on the A-site.
Overall charge neutrality is automatically maintained by this expression.
With this definition the configurationally dependent electrostatic energy
now depends on both the A- and B-sublattices:
\begin{equation}
 E_{\rm AB}={e^2\over2 \epsilon} 
  \left [ 
   \sum_{(lA)\ne(l'A')} 
      {\Delta \widetilde{q}_{lA}\Delta \widetilde{q}_{l'A'}
      \over\, |{\bf R}_{lA}-{\bf R}_{l'A'}|} 
   + 
   \sum_{(lB)\ne(l'B')} {\Delta q_{lB}\Delta q_{l'B'}
      \over\, |{\bf R}_{lB}-{\bf R}_{l'B'}|} 
   + 
   2 \sum_{lA,l'B} {\Delta \widetilde{q}_{lA}\Delta q_{l'B}
    \over\, |{\bf R}_{lA}-{\bf R}_{l'B}|} 
  \right ] \;,
\label{cteq}
\end{equation}
This expression is formally similar to Eq. (\ref{ABeq}), except that
here the charge on an A-site depends not only on the atomic species on
that site but on the nearest neighbor B-site configuration.
 
The value of the additional parameter $\lambda$ can be chosen
depending on the species occupying the A-site. Thus for Ba-based
alloys $\lambda$~=~0, no charge transfer occurs, and the model reduces
to Bellaiche and Vanderbilt's model. For Pb-based alloys, the
value of $\lambda$ can be fit to first-principles calculations of
representative supercells of Pb(BB')O$_3$.

The parameters in our model are chosen as follows.  $\epsilon$ is
chosen to reproduce the energy differences of selected Ba-based alloys
(we used $\epsilon$~=~10, as in Ref. \cite{bc}). The value of
$\lambda$ is then fit to selected Pb-based alloy energy differences.
Fig. 1 shows structural energy differences ($x$~=~0) as a
function of $\lambda$. To obtain B-site stacking structural energy differences
in the range of
10 kJ/(mol-ABO$_3$) for Pb based alloys
energies, as found in first-principles calculations, Fig. 1 indicates
that $\lambda_{\rm Pb}$ should be chosen in the range $0.7 < \lambda_{\rm Pb}$.

\section{Monte Carlo Simulations}

Metropolis Monte Carlo simulations\cite{Metropolis} 
were carried out with periodic boundary conditions
using 6$\times$6$\times$6 supercells, similar to those in BV.
To obtain good statistics, typically $\sim 10^7$ moves were made at
each temperature. Calculations were started at at high temperature
(e.g., 4000~K) with a random configuration, and the temperature was
then slowly decreased until the acceptance rate becomes very low.
To determine the B-site ordering the Fourier
transform of the charge-charge correlation function $\eta({\bf k})$ was
computed from converged runs at a given temperature: \cite{bv}
\begin{equation}
 \eta({\bf k}) = \alpha \sum_{ll^{'}}
     \Delta q_l \, \Delta q_{l+l^{'}}
     \, \exp(-i \bf k\cdot l^{'}) \;,
\end{equation}
where $\alpha$ is a normalization factor, the sum runs over
the B-sublattice, and $\bf k$ is the wavevector
in the Brillouin zone of the unit cubic cell.
The charge-charge correlation function $\eta({\bf k})$ is directly 
proportional to the ensemble average of the square of 
the Fourier transform of the charge distribution, 
e.g., a large value of $\eta$ at
${\bf k}=2\pi({1\over2},{1\over2},{1\over2})$ or
${\bf k}=2\pi({1\over3},{1\over3},{1\over3})$
corresponds to a strong [111]$_{1:1}$ or [111]$_{1:2}$ order
respectively.

\section{Results}

MC calculations using
electrostatic models with and without charge transfer are reported in
this section as a function of temperature and concentration
$x$ of A- and B-site doping. 
Results for B-site only alloys are presented first.
A- and B-site alloy calculations are then considered, first using
a purely electrostatic models and then 
including charge transfer.

\subsection{B-site Alloys With Charge Transfer}

Simulations were performed using Eqs. (\ref{lambdaeq},\ref{cteq}) for
A(BB$^{'}$B$^{''}$)O$_3$ alloys with $\lambda$~=~0.3 and 0.7, and
setting $a$~=~7.7~a.u. and $\epsilon$~=~10. Results for
II$_{(1-x)/3}$IV$_x$V$_{2(1-x)/3}$ heterovalent ternaries are given in
Fig. 2, which shows the calculated charge-charge correlation function
$\eta({\bf k})$ vs.\ tetravalent composition $x$ at temperatures of
$T$~=~1000~K and 1500K and for $\lambda$~=~0.3 and 0.7
respectively. For $\lambda$~=~0.3, which permits only a small charge
transfer between A-sites, Bellaiche and Vanderbilt's prediction is
reproduced (also for $\lambda$~=~0.0, not shown), {\it i.e.} we obtain the
B-site ordering sequence of Ba-based perovskite alloys described in
Section II.  For $\lambda$~=~0.7, Fig. 2 shows there is no
[111]$_{1:2}$ ordered phase.  Instead, there is a continuous
transformation from [111]$_{1:1}$ order at $x$~=~0 to a disordered
state at large $x$. Increasing the final temperature to T=1500K does
not change this ordering sequence, but only decreases the value of $x$
at which disorder appears, as expected.

Moreover, the [111]$_{1:1}$ structure is similar to the
``random-site'' model, \cite{ad} in that it has alternating planes of
mixed (II-V)-planes and pure (V)-planes, as was also found by BV.
These results are in good agreement with experimental observations of
the ordering sequence of Pb(Mg$_{1/3}$Nb$_{2/3}$)O$_{3}$ (PMN),
\cite{Husson1,Husson2,Chen}
(1-$x$)~Pb(Mg$_{1/3}$Ta$_{2/3}$)O$_{3}$~+~$x$~PbZrO$_{3}$ (PMT-PZ)
\cite{ad2} and Pb(Mg$_{1/3}$Ta$_{2/3}$)O$_{3}$ (PMT). \cite{ad3}

The charge transfer model also correctly predicts
B-site ordering in the III$_{(1-x)/2}$IV$_x$V$_{(1-x)/2}$ and
II$_{(1-x)/2}$IV$_x$VI$_{(1-x)/2}$ alloys. Monte Carlo simulations
using Eqs. (\ref{lambdaeq},\ref{cteq}) with $\lambda$~=~0 and 
$\lambda$~=~0.7 
show that both Ba-based and Pb-based perovskite alloys
have the same B-site ordering for III$_{(1-x)/2}$IV$_x$V$_{(1-x)/2}$
ternaries and II$_{(1-x)/2}$IV$_x$VI$_{(1-x)/2}$ ternaries. Our results
reproduce those of BV in these cases.

\subsection{A- and B-site Alloys Without Charge Transfer}

To study the effects of A-site chemical substitution, we first used a
purely electrostatic model with formal ionic charges. Calculations
were performed for
(A$_{1/2}^{1+}$A$^{'~3+}_{1/2}$)(B$_{1/3}^{2+}$B$^{'~5+}_{2/3}$)O$_3$
and
(A$_{1-x}^{2+}$A$^{'~3+}_{x}$)(B$_{(1+x)/3}^{2+}$B$^{'~5+}_{(2-x)/3}$)O$_3$.
Since $Q_{l,\rm A}$ is now not a constant, we express the A- and B-sublattice
charges using Eq. (\ref{ABeq}). 

Fig. 3 shows both A- and B-site charge-charge correlation function
$\eta({\bf k})$ vs.\ temperature for
(A$_{1/2}^{1+}$A$^{'~3+}_{1/2}$)(B$_{1/3}^{2+}$B$^{'~5+}_{2/3}$)O$_3$.
The purely electrostatic model, Eq. (\ref{ABeq}), can
explain the observed [111]1:1 B-site ordering of
(Na$_{1/2}$La$_{1/2}$)(Mg$_{1/3}$Nb$_{2/3}$)O$_3$. \cite{dcd} Moreover, 
the 1:1 B-site ordering found in the calculations
corresponds to the random site model. 
As seen in Fig. 3, weak [001]$_{1:1}$ A-site ordering is also
obtained in the simulations, in agreement with measurements
on samples annealed below $\sim 950^{\circ}$C \cite{dcd}.
Perfect [001]$_{1:1}$ A-site ordering would
correspond to alternating [001] A-planes of Na and La, but the
A-sites in the simulations are more disordered (see section VI). 
In simulations using a 6$\times$6$\times$6 supercell, only B-site order was
observed. The weak A-site ordering emerged only after increasing the 
simulation region to a larger 12$\times$12$\times$12 supercell.

Fig. 4 shows $\eta({\bf k})$ vs.\ tetravalent composition $x$ for
(A$_{1-x}^{2+}$A$^{'~3+}_{x}$)(B$_{(1+x)/3}^{2+}$B$^{'~5+}_{(2-x)/3}$)O$_3$.
The symmetry about $x~=~0.5$ is a consequence of our model. Exchanging
$x$ with $1-x$ in Eq. (\ref{ABeq}), every $\Delta q_{l\tau}$ changes
sign, and the configurational energy is unchanged. 
The strongest [111]1:1 B-site order occurs at $x$~=~0.5, in agreement with
experiment.
Experimentally,
(Ba$_{1-x}$La$_{x}$)(Zn$_{(1+x)/3}$Nb$_{(2-x)/3}$)O$_3$ (BLZN)
exhibits 1:2 ordering coexisting with 1:1 B-site order for $x$ less
that $\sim 0.05$, then switching to 1:1 B-site order for $0.05 < x <
0.6$. \cite{ad5}
(Ba$_{1-x}$La$_{x}$)(Mg$_{(1+x)/3}$Nb$_{(2-x)/3}$)O$_3$ (BLMN) behaves
similarly except that 1:1 B-site order is observed over the whole
range $0.05 < x < 1$. \cite{ad4} 
Fig. 4 shows that the purely electrostatic
model, Eq. (\ref{ABeq}), correctly reproduces the observed B-site
ordering, except for the extreme La-rich $x > 0.95$ region, where
Eq. (\ref{ABeq}) incorrectly predicts 1:2 ordering. However, including
charge transfer (see below) improves the agreement with experiment in
the extreme La-rich region (section C below).

The character of the 1:1 B-site order in Fig. 4 changes at $x = 1/2$.
A detailed examination of the equilibrated Monte Carlo structures
shows that the 1:1 order results from a [111] rocksalt-like 
alternation of $\beta'$ and $\beta''$ layers. For $x<0.5$ we find
$[\beta']_{1/2}[\beta'']_{1/2} = [(\rm B^{2+}_{(2+2x)/3})(\rm
B^{5+}_{(1-2x)/3})]_{1/2}[\rm B^{5+}]_{1/2}$, {\it i.e.} 
the $\beta'$ site
is a mixture of B$^{+2}$ and B$^{+5}$ atoms, and the $\beta''$ site is
solely occupied by B$^{+5}$ atoms. This is consistent with the
``random-site'' model. 
The peak in the 1:1 order parameter at $x=0.5$ in Fig. 4 corresponds
to $[\beta']_{1/2} = [\rm B^{2+}]_{1/2}$, so the $\beta'$ and $\beta''$ are
each occupied by a single cation species, and the 1:1 reflections \cite{ad4}
exhibit their maximum intensity.
However, for $0.5<x$ we find
$[\beta']_{1/2}[\beta'']_{1/2} = [\rm B^{2+}]_{1/2}[(\rm
B^{2+}_{(2x-1)/3})(\rm B^{5+}_{(4-2x)/3})]_{1/2}$, where now the
$\beta'$ site is solely occupied by B$^{+2}$ atoms and the $\beta''$
site is now a mixture of B$^{+2}$ and B$^{+5}$ atoms.  Akbas
{it et al.} \cite{ad5,ad4,mad} proposed this structure based on
their observations.

\subsection{A- and B-site Alloys With Charge Transfer}

We next consider whether the inclusion of charge transfer may improve the
agreement with experiment for the A- and B-site alloys considered in
the previous section. The electronegativity and electron affinity of
an atom is an indicator of the degree of covalency. By this measure,
La is more covalent than Ba or Na, and Nb has higher covalency than
Ta.

There is a difficulty in generalizing our charge transfer model to
systems with more than one atomic species on the A- and B-sites, such as
(Ba$_{1-x}$La$_{x}$)(Mg$_{(1+x)/3}$Nb$_{(2-x)/3}$)O$_3$. Considering
charge transfer only onto La ions and using Eq. (\ref{lambdaeq}), charge
neutrality may be violated except for $x$~=~1. For BLMN, the simplest way to
restore charge neutrality is to permit charge transfer onto both La
and Ba ions, using an average charge trasfer parameter $\lambda_{\rm A}(x)$ proportional to the La composition
$x$, i.e., $\lambda_{\rm A}(x)=x\lambda_{\rm A}(1)$,
so that there is no charge transfer at $x$~=~0 (BMN) and the largest
charge transfer occurs for at $x$~=~1 (LMN). 

In BLMN alloys, charge transfer onto B-sites may also be considered.
Mg is pure ionic, but Nb and Ta are more polarizable.
To ensure charge neutrality we may define
$\lambda_{\rm B}$ similar to $\lambda_{\rm A}$: 
\begin{equation}
 \lambda_{\rm B}(x)={(2-x)\over 3} \lambda_{\rm A}(1) \;,
\end{equation}
which varies linearly with $B^{+5}$ concentration.
The configurationally dependent charges on the A-sites and B-sites
are then defined as:
\begin{equation}
 \Delta \widetilde{q}_{l,\rm A}=-\lambda_{\rm A}(x)({1\over8} \sum_{i=n.n. {\rm B-sites}}
                 {\Delta q_{i,\rm B} }) \;,
\end{equation}
\begin{equation}
 \Delta \widetilde{q}_{l,\rm B}=-\lambda_{\rm B}(x)({1\over8} \sum_{i=n.n. {\rm A-sites}}
                 {\Delta q_{i,\rm A} }) \;,
\end{equation}
where
\begin{equation}
 \Delta q_{l,\tau}=Q_{l,\tau}-q_{\tau} \;,
\end{equation}
where, $\tau$~=~\{A,B\}, $Q_{l,\tau}$ is the formal charge, and $q_{\rm A}$ and $q_{\rm B}$ 
are the mean charge of A-sites and B-sites respectively. 

We first consider BLMN. To test this model, we set $\lambda_{\rm
A}(1)$~=~0.7, since LMN is observed to have [111]$_{1:1}$ B-site
ordering.  Fig. 5 shows $\eta({\bf k})$ as a function of La
composition $x$ for BLMN. The main effect of including charge transfer
is that 1:1 order now appears in the La-rich region. 
There is
essentially no difference in the B-site ordering with and without
charge transfer onto the B-sites. Both display [111]$_{1:2}$ order at
small $x$ ($x<3\%$), then [111]$_{1:2}$ and [111]$_{1:1}$ order
coexist at $3\%<x<5\%$, and finally [111]$_{1:1}$ order for all
$x>5\%$. 
However, the model introduces
features not seen in the measurements of BLMN. \cite{ad4} The peak of
[111]$_{1:1}$ order shifts from $x = 0.5$ to $x\sim 0.35$, and there
is a dip in $\eta({\bf k})$ for [111]$_{1:1}$ order at $x\sim 0.8$.

The replacement of Ba by Pb in the La-doped perovskites
(Pb$_{1-x}$La$_{x}$) (B$_{(1+x)/3}^{2+}$B$^{'~5+}_{(2-x)/3}$) O$_3$
is now considered.
For PLMN, 1:1 B-site ordering is observed for La concentrations,
increasing continuously up to $x=0.5$ and the decreasing for $x >
0.5$. Again, this trend is consistent with Fig. 4, except that the purely
electrostatic model results in 1:2 order in the extreme Pb and La rich 
regions. \cite{mad}
Several choices for the charge transfer parameter were considered.
Assuming first that La is much less covalent than Pb,
the charge transfer constant is chosen to be proportional to the Pb
concentration, $\lambda_{\rm A}(x)=(1-x)\lambda_{\rm A}(1)$,
so that there is no charge transfer at $x$~=~1 (LMN) and the largest
charge transfer occurs for at $x$~=~0 (PMN). The resulting B-site ordering
would be similar to Fig. 5, but with $x$ mapped to $1-x$. Compared to Fig. 4,
this improves the agreement with experiment for the Pb-rich alloys,
yielding 1:1 B-site order. However,
for La concentrations greater than about 95\% the model does not yield 
the observed 1:1 ordering.\cite{mad}
On the other hand assuming that La and Pb are similarly covalent,
$\lambda_{\rm A}$ is set to a constant, $\lambda_{\rm A} = 0.7$
as in PMN. The results are shown in Fig. 6. This model agrees
with experiment in the Pb-rich and La-rich regions, reproducing the
observed 1:1 B-site ordering, but is disordered for
intermediate concentrations. Including charge
transfer onto the B-sites with $\lambda_{\rm B}(x)=0.7{(2-x)\over 3}$ 
(not shown) yields similar results to those with charge
transfer on A-sites only. 

\section{Discussion}

First-principles calculations show that the energy differences between
different B-site stackings of Pb-based
\cite{bc,wensell99,wensell-2000} alloys are greatly reduced compared
to those of Ba-based \cite{bc} alloys. For example, the energy
differences between [100]1:2, [110]1:2, [111]1:2 B-site stackings are
reduced from $\sim$ 60~kJ/(mol-ABO$_3$) in Ba-based alloys to $\sim$
5~kJ/(mol-ABO$_3$) in Pb-based alloys. \cite{bc} The energy reduction
can be attributed to the greater covalency of Pb compared to Ba, which
results in larger local relaxation energies in Pb-based compared to
Ba-based alloys. As shown in Fig. 1, there is a dramatic decrease in
structural energy differences when charge transfer between
A-sites is included.

The magnitude of local relaxation energies due to
Pb's greater covalency but {\it
not including} charge transfer between A-sites can be estimated as
follows. The ferroelectric PbTiO$_3$ double-well has a depth
of $\sim$ 10 kJ/(mol-ABO$_3$) not
including strain, which is much larger than the well depth in
BaTiO$_3$, consistent with the more covalent nature of
Pb. \cite{cohen_krakauer_92} However 10 kJ/(mol-ABO$_3$) is still not enough
to account for the reduction from the 60 kJ/(mol-ABO$_3$) B-site stacking 
energy differences in Ba-based alloys. In 
heterovalent B-site alloys, however, charge transfer between A-sites is
another possible local relaxation mechanism. 

This is especially true
for small $x$ in B-site ternaries like II$_{(1-x)/3}$IV$_x$V$_{2(1-x)/3}$.
Since the A-site has a nearest neighbor shell of eight B-sites, each
of which is occupied by a $\Delta q_{l,\rm B}=-2$ or 
$\Delta q_{l,\rm B}=+1$  ion (for $x = 0$),
the sum of these eight $\Delta
q_{l,\rm B}$ will necessarily be non-zero in the $x = 0$ limit. 
By contrast, in III$_{(1-x)/2}$IV$_x$V$_{(1-x)/2}$
and II$_{(1-x)/2}$IV$_x$VI$_{(1-x)/2}$ alloys this ``non-neutrality''
condition is not forced.
Our model simulates
the effective charge transfer between A-sites that results
from a combination of oxygen and Pb atomic relaxations and direct
electronic charge transfer. The purely ionic character of Ba inhibits
this, but the greater covalency of Pb makes this an effective
mechanism to increase the local relaxation energy. 

Local relaxation in Pb based alloys plays an important role in
compensating the purely electrostatic energy differences arising from
different B-site stackings. The electrostatic charge transfer model
in Section III correctly describes the observed B-site ordering in PMN
type B-site II$_{(1-x)/3}$IV$_x$V$_{2(1-x)/3}$ heterovalent
ternaries. As shown by Fig. 2, introducing charge transfer changes the
small $x$ ordering from [111]1:2 to the [111]1:1 ordering seen
experimentally.

While improving agreement with experiment for II-IV-V B-site alloys,
the inclusion of charge transfer does not alter the correct B-site
ordering predictions of the BV model for the III-IV-V and II-IV-VI
alloys. For these alloys, the simulations predict a rocksalt-like 
structure where the
A-site has on average a ``neutral'' nearest neighbor shell of
B-sites. In this case, the average $\Delta \widetilde{q}_{\rm A}$ will be
small for any value of $\lambda$. 
Direct Monte Carlo calculations using our charge transfer model
predict the same B-site ordering as the purely electrostatic model of 
Bellaiche and Vanderbilt.

A detailed analysis of our results supports the random-site model
of the [111]1:1 ordered regions. We do not find any evidence for
the space-charge model. In the
space-charge model, as discussed in Ref. \cite{ad}, the 1:1 ordered
regions correspond to a rock-salt like [111] stacking of alternating
II:V planes, which is locally charge-imbalanced. 
In the space-charge scenario, Coulomb repulsion would limit the 
size of these charge-imbalanced regions to nanoscale size. 
Overall charge neutrality is restored
by embedding the nanoscale 1:1 regions 
in a disordered (+5 ion)-rich matrix. The space-charge model was
long favored, because it could explain the failure of 
high temperature annealing to increase the size of these 1:1 regions.
Instead, the recent successful coarsening of these 1:1 regions \cite{ad}
would seem to rule out charge-imbalanced 1:1 regions. 
The random-site model is also supported by recent direct Z-contrast 
imaging in PMN. \cite{xu2000}

Indeed, our calculations show that the [111]1:1 ordering
corresponds to alternating (II-V):V planes, {\it i.e.}  planes
with a random mixture of (II-V) cations alternating with pure V planes.
Bellaiche and Vanderbilt also observed this structure in their 1:1
ordered regions. 
This stacking is consistent with the random-site
model. In materials that are also A-site alloys, our calculations 
reveal interesting
variants of the random-site model, as discussed below.

The results of A- and B-site alloy B-site ordering presented
in the previous section show that the purely electrostatic model,
Eq. (\ref{ABeq}), can account for many of the observations. 
For (Na$_{1/2}$La$_{1/2}$)(Mg$_{1/3}$Nb$_{2/3}$)O$_3$ Dupont
{\it et al.} \cite{dcd} argued that [111]1:2 B-site order in
is incompatible with
the different valences of the A-site cations. In the [111]1:2
structure there are two types of environments for the A-site.
In the first, two A-sites are located between a B$^{5+}$ and a B$^{2+}$
layer and have 5 B$^{5+}$ and 3 B$^{2+}$ nearest neighbors. The formal
charge on these subcells is $[(5(+5)+3(+2))/8-6] = -2.125$. The second
A-site is located between two B$^{5+}$ layers and has 6 B$^{5+}$ and 2
B$^{2+}$ nearest neighbors, with a subcell formal charge of
-1.75. With a 1:1 mixture of Na$^{+1}$ and La$^{+3}$ on the A-sites,
each cation must occupy both types of A-site, leading to local charge
imbalances of -1.125 and -0.75 for Na and +0.875 and +1.25 for La on
the two respective type of A-sites. These are much larger than the
charge imbalances when all A-site are occupied by divalent cations,
namely -0.125 and +0.25, respectively.
By contrast, in the random-site 1:1 B-site order structure,
all the A-sites are equivalent and have 4 B$^{5+}$ and 4
(B$^{5+}_{2/3}$ B$^{5+}_{1/3}$) neighbors, with a formal subcell
charge of -2. This leads to local charge imbalances of -1 for Na and
+1 for La. Dupont {\it et al.} noted that this reduction could be a factor
in promoting the [111]1:1 B-site ordered structure. The MC results in
Fig. 3, using the electrostatic-only model, confirm that 1:1 order can arise
from purely electrostatic interactions. 

As mentioned, the weak A-site ordering was not seen in simulations
using a 6$\times$6$\times$6 supercell, but only in the larger
12$\times$12$\times$12 supercell shown. 
The large error bars in Fig. 3 reflect the difficulty in equilibrating
this system at the lower temperatures in the Monte Carlo runs.
This is probably due to the presence of many structural states
with nearly degenerate energies.
Perfect [001]$_{1:1}$ A-site ordering would correspond
to alternating [001] A-planes of Na and La, but this is not seen
in the simulations and is reflected in the small value of the order parameter.
To test if perfect [001]$_{1:1}$ A-site ordering is a global energetic
minimum, we constrained the A-sites to have this ordering 
and allowed only the B-site configuration to change. 
This resulted in much weaker [111]1:1 B-site order than seen in Fig. 3
and indicates the complex nature of the A-site order.

In BLMN, Fig. 4 shows that purely electrostatic interactions alone can
account for stabilization of 1:2 B-site ordering for small La
concentrations, switching to 1:1 ordering for larger
concentrations. 
The experimentally observed 1:1 reflections \cite{ad4} also 
have a peak at $x=0.5$, which is consistent with the random-site 
model. For $x < 0.5$ we find the $\beta'$ site is occupied by a random mixture
of B$^{+2}$ and B$^{+5}$ atoms, while for $x > 0.5$ the $\beta''$ site 
is occupied by a random mixture. At $x=0.5$, the $\beta'$ and $\beta''$ are
each occupied by a single cation species, yielding a peak in the 1:1 
reflections.
Experimentally, the 1:1 ordering persists up to the
extreme La rich region \cite{ad4}. However, the intensity of the 1:1
ordering reflections decreases between $0.5 < x < 1.0$, as 
reproduced by Fig. 4. 
Akbas and Davies have argued that the
1:2 structure structure is stabilized by anion displacements. 
In BMN, 1:2 MNN ordering into
separate B$^{+2}$ and B$^{+5}$ layers permits adjacent anion
layers to undergo a concerted displacement towards the smaller
B$^{+5}$ ion. This requires Nb cations to have an asymmetric anion
coordination. For $x=1$ (LMN), 1:2 MMN ordering would lead to an asymmetric
environment for Mg, which is rarely observed. \cite{ad4} 
Akbas and Davies argued
that this stabilizes the 1:1 ordered phase for $x=1$. 
The present calculations show that
purely electrostatic interactions alone stabilize the 1:2
ordering near $x=0$ and $x=1$, since atomic displacements are
not included.
The results in Fig. 5 and 6 show that 1:1 order at
$x=1$ can be stabilized by the inclusion of charge transfer, which
serves to reduce the energy differences between different B-site orderings,
as shown in Fig. 1. For BLMN and PLMN, we interpret this charge transfer 
as modeling 
larger local structural relaxations for La than for Ba, tipping the balance
slightly in favor of 1:1 ordering.

\section{Conclusions}

We have used electrostatic models with
and without charge transfer to study
B-site ordering in materials
with alloying on both the A- and B-sites. 
For Pb$^{2+}$ and possibly other covalent ions, 
the effective charge on the A-site can depend on the local
B-site configuration.
The inclusion of charge transfer into the
electrostatic model of Bellaiche and Vanderbilt is
shown to reproduce the observed B-site ordering 
in PMN type ternary perovskite alloys. 
For A- and B-site alloys, general trends are reproduced by purely
electrostatic interactions. 
In (Na$_{1/2}$La$_{1/2}$)(Mg$_{1/3}$Nb$_{2/3}$)O$_3$, electrostatic
effects alone produce B-site and A-site ordering observed in
experiments.
Charge transfer 
effects in simulations for Pb-based A- and B-site
alloys indicate that local structural
relaxations can tip the balance between different B-site
orderings in some cases.
With the inclusion of charge transfer in these cases, 
we conclude that the electrostatic interaction is the leading mechanism 
responsible for the B-site ordering in perovskite alloys.

\acknowledgements

This work is supported by the Office of Naval Research grant
N00014-97-1-0047. We acknowledge useful discussions with Peter Davies
and Shiwei Zhang. We are grateful to Suhuai Wei for discussions on
electrostatic models incorporating charge transfer.


\newpage

\begin{figure}
\epsfxsize=3.2in
\epsfysize=2.5in
\ \leftline{\epsfbox{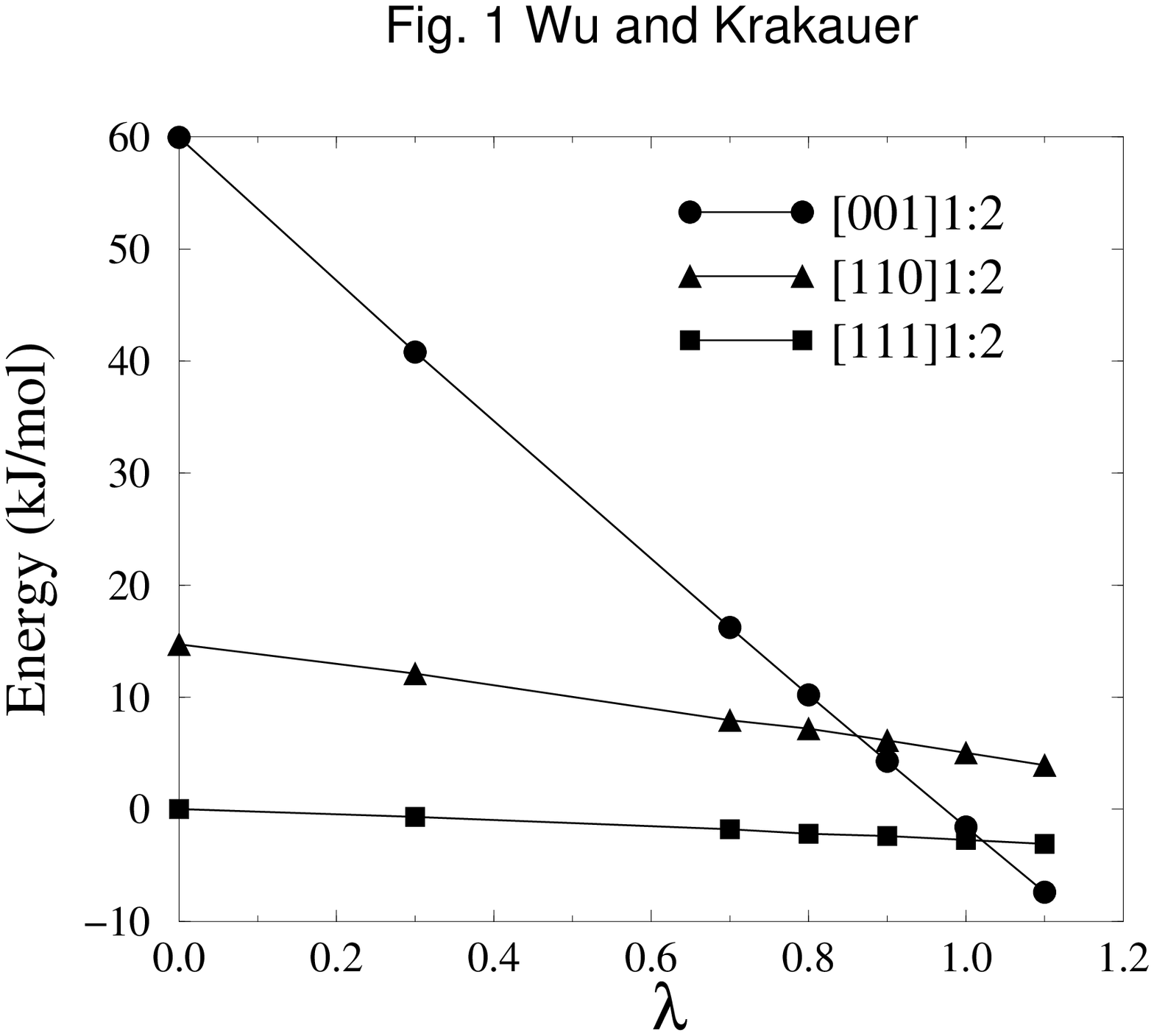}}
\caption{Total energies (per mol = ABO$_3$) for 15-atom supercell 1:2
B-site stackings of A(B$_{1/3}$B$_{2/3}^{'}$)O$_{3}$.  Energies are
relative to the $\lambda = 0$ value of $[111]_{1:2}$.}
\vskip0.05in
\label{fig-E-lamb}
\end{figure}

\begin{figure}
\epsfxsize=3.2in
\epsfysize=3.0in
\ \leftline{\epsfbox{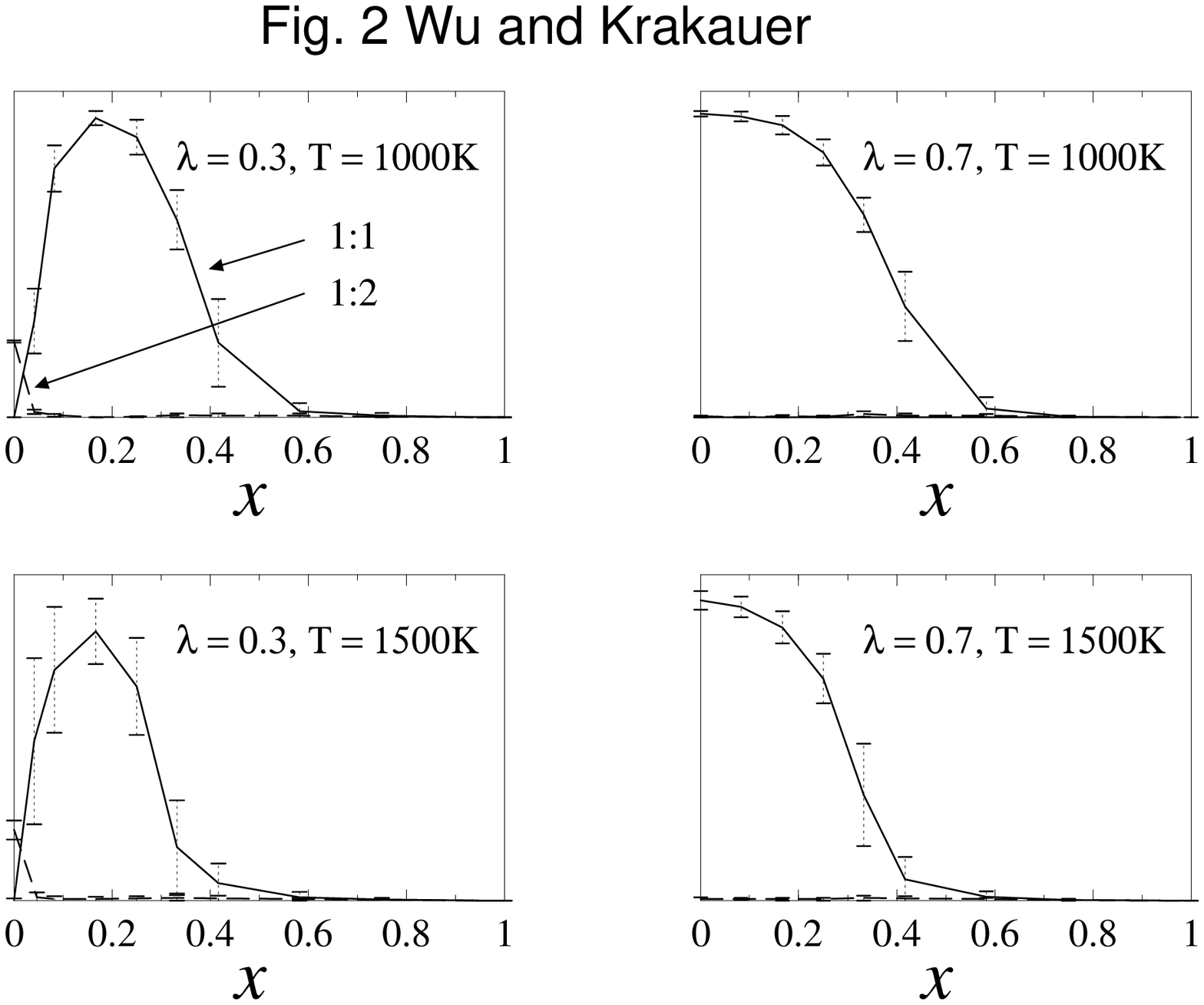}}
\caption{Monte Carlo simulations of the long-range order parameter
 $\eta({\bf k})$ {\it vs.} tetravalent atomic composition $x$
 using the charge transfer model for II$_{(1-x)/3}$IV$_x$V$_{2(1-x)/3}$ 
 ternaries, with 6$\times$6$\times$6 supercells, $a$~=~7.7~a.u., 
 and $\epsilon$~=~10.}
\vskip0.05in
\label{fig-eta-Bsite}
\end{figure}

\begin{figure}
\epsfxsize=3.2in
\epsfysize=2.9in
\ \leftline{\epsfbox{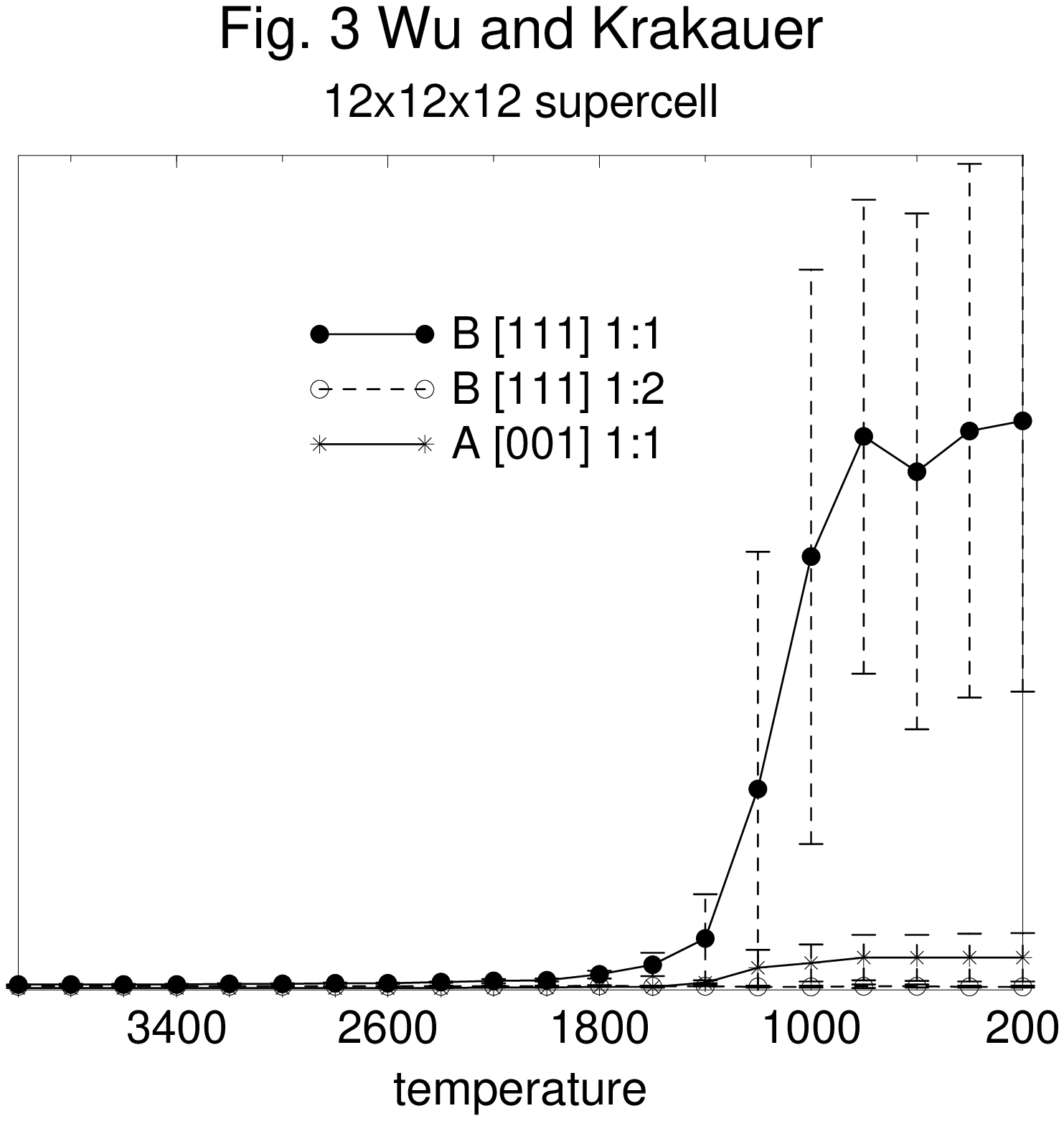}}
\caption{Monte Carlo simulations of the long-range order parameters
 $\eta({\bf k})$ {\it vs.} temperature T using BV model for 
 (A$_{1/2}^{1+}$A$^{'~3+}_{1/2}$)(B$_{1/3}^{2+}$B$^{'~5+}_{2/3}$)O$_3$  
 perovskites, with 12$\times$12$\times$12 supercells, $a$~=~7.5~a.u., 
 and $\epsilon$~=~10.}
\vskip0.05in
\label{fig-eta-NaLa}
\end{figure}

\begin{figure}
\epsfxsize=3.2in
\epsfysize=2.9in
\ \leftline{\epsfbox{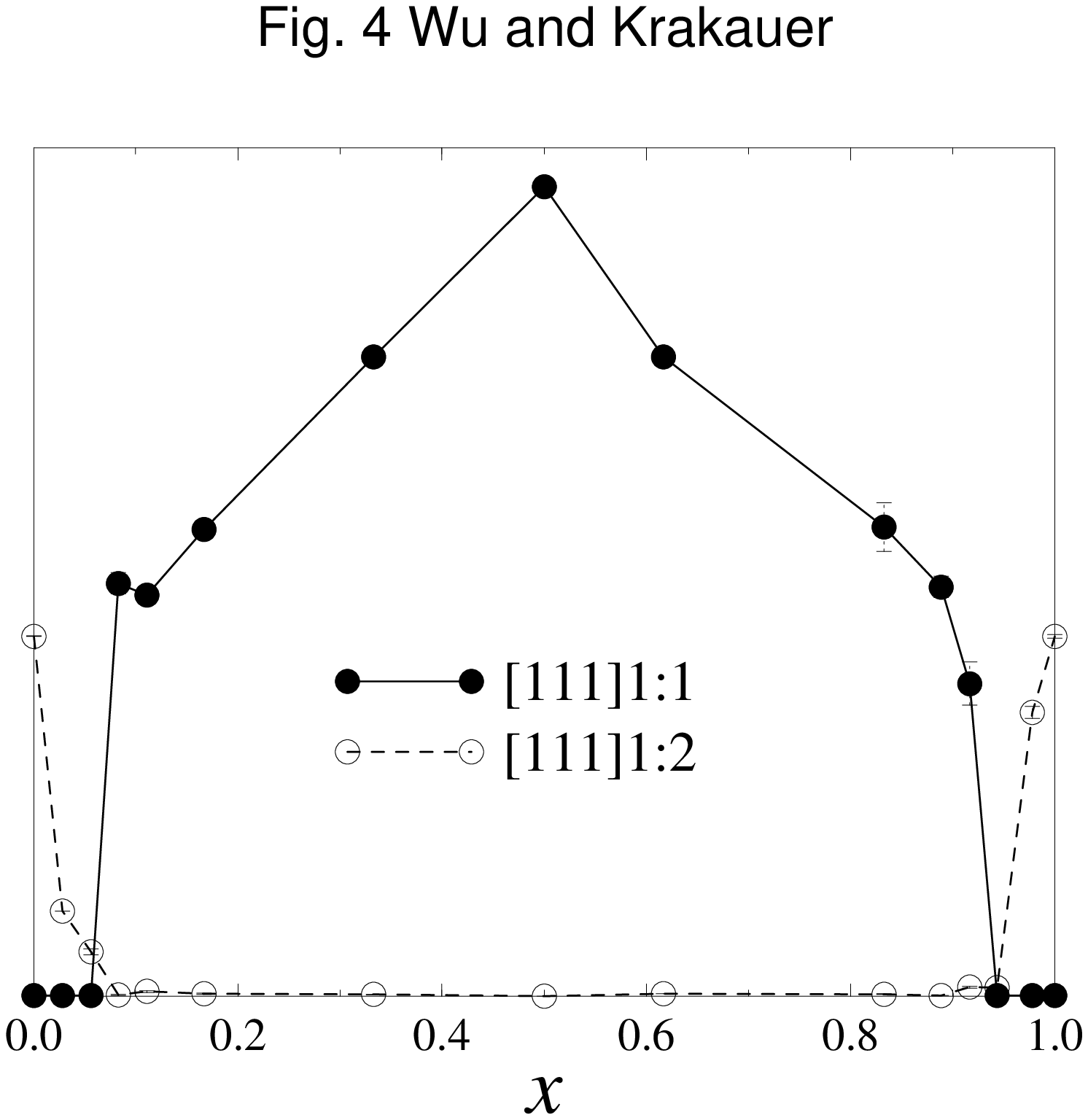}}
\caption{Monte Carlo simulations of the long-range order parameter
 $\eta({\bf k})$ {\it vs.} A$^{'~3+}_{x}$ composition $x$ using BV model for
 (A$_{1-x}^{2+}$A$^{'~3+}_{x}$)(B$_{(1+x)/3}^{2+}$B$^{'~5+}_{(2-x)/3}$)O$_3$
 perovskites, with 6$\times$6$\times$6 supercells, $a$~=~7.5~a.u., 
 and $\epsilon$~=~10, at temperature T = 1000~K.}
\vskip0.05in
\label{fig4}
\end{figure}

\begin{figure}
\epsfxsize=3.2in
\epsfysize=2.9in
\ \leftline{\epsfbox{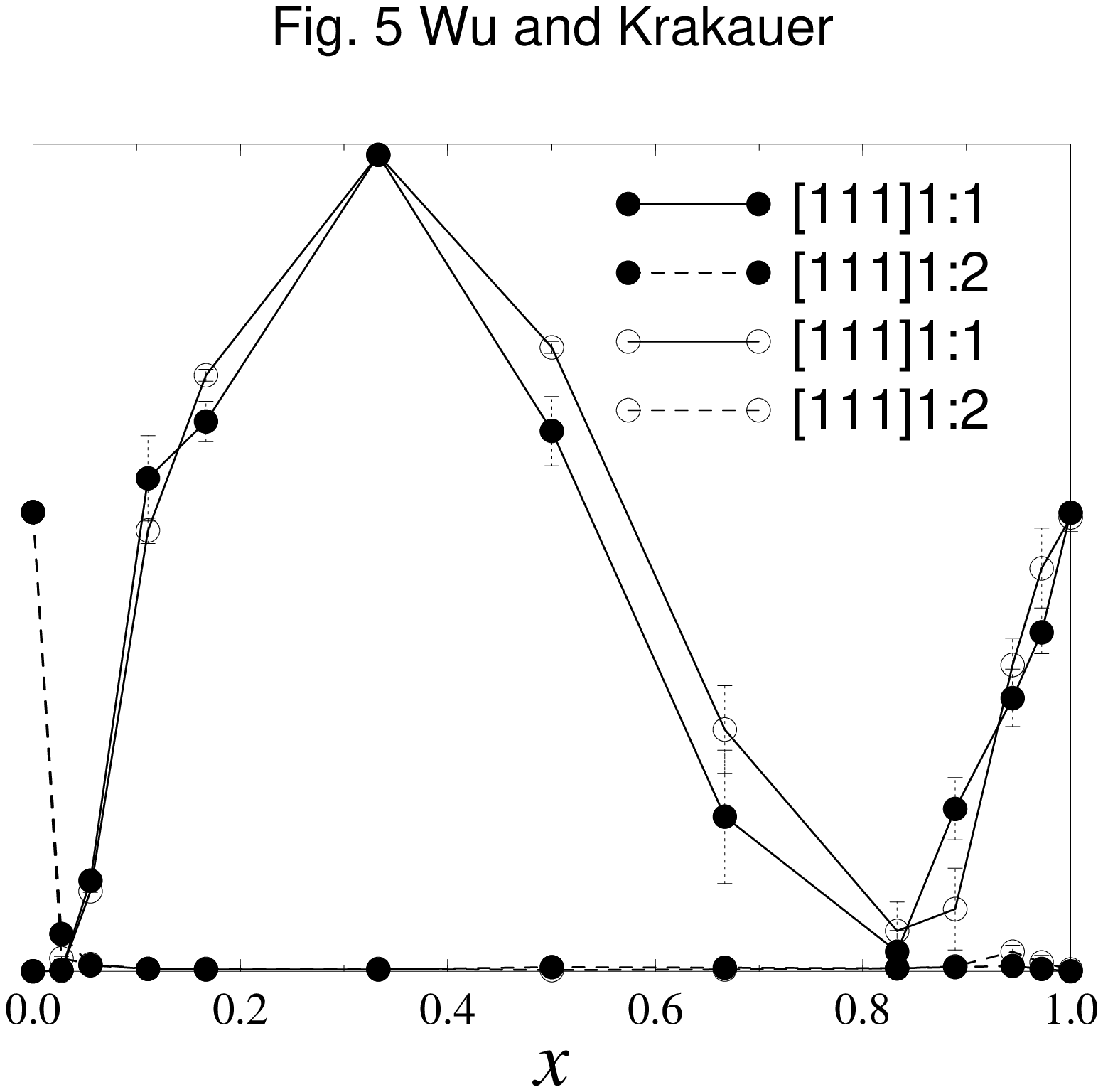}}
\caption{Monte Carlo simulations of the long-range order parameter
 $\eta({\bf k})$ {\it vs.} La composition $x$ using charge transfer 
 model for 
 (Ba$_{1-x}$La$_{x}$)(B$_{(1+x)/3}^{2+}$B$^{'~5+}_{(2-x)/3}$)O$_3$  
 perovskites, with 6$\times$6$\times$6 supercells, $a$~=~7.5~a.u., 
 and $\epsilon$~=~10, at temperature T = 1000~K,
 $\lambda_{\rm A}(x)=0.7x$, $\lambda_{\rm B}(x)=0.7(2-x)/3$.
 The solid symbols denote
 charge transfer only on A-sites, and the unfilled symbols denote
 charge transfer on both A-sites and B-sites.}
\vskip0.05in
\label{fig5}
\end{figure}

\begin{figure}
\epsfxsize=3.2in
\epsfysize=2.9in
\ \leftline{\epsfbox{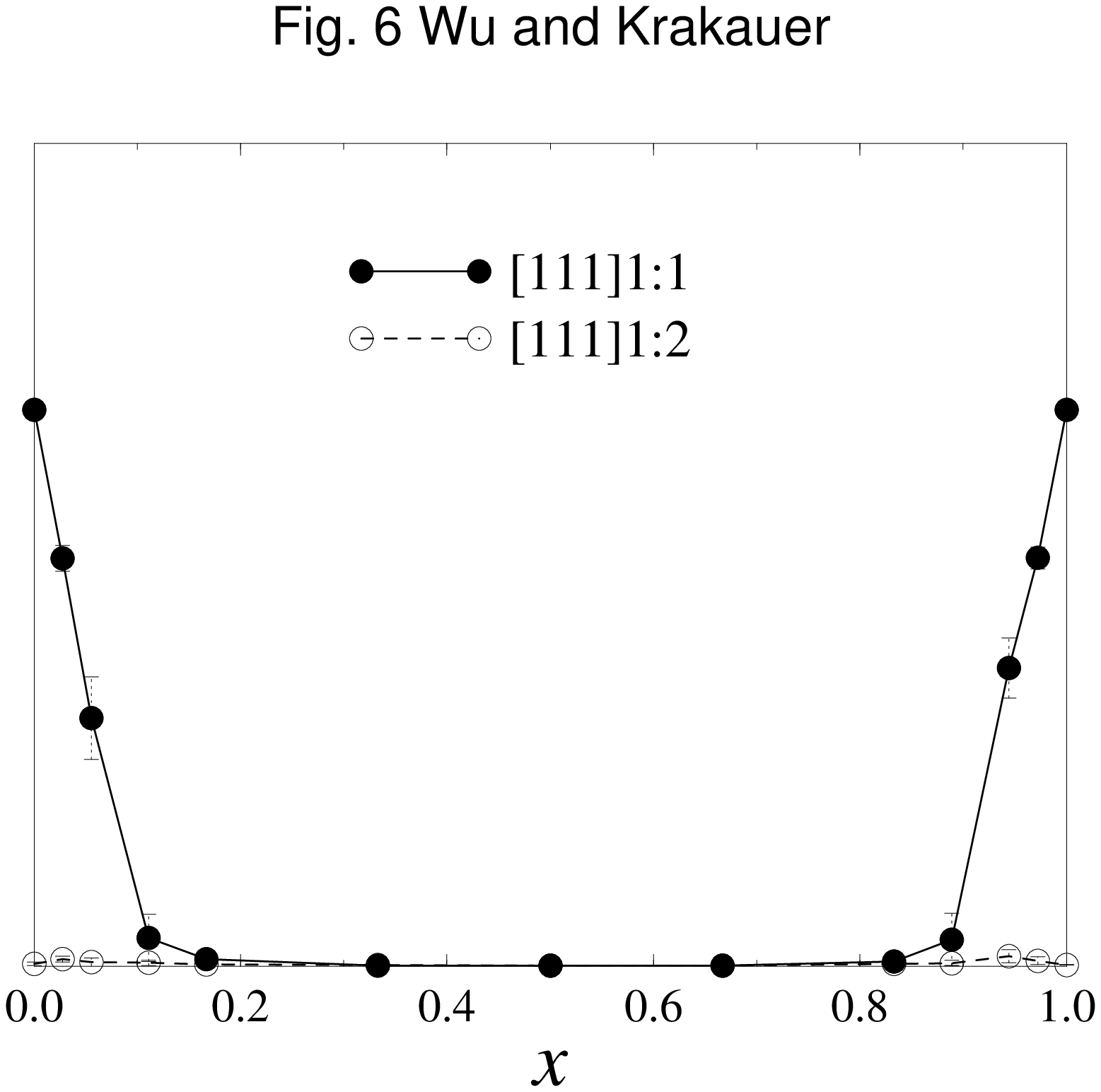}}
\caption{Monte Carlo simulations of the long-range order parameter
 $\eta({\bf k})$ {\it vs.} La composition $x$ using charge transfer 
 model only on A-sites for
 (Pb$_{1-x}$La$_{x}$)(B$_{(1+x)/3}^{2+}$B$^{'~5+}_{(2-x)/3}$)O$_3$
 perovskites, with 6$\times$6$\times$6 supercells, $a$~=~7.5~a.u., 
 and $\epsilon$~=~10, at temperature T = 1000K. $\lambda_{\rm A}=0.7$.}
\vskip0.05in
\label{fig6}
\end{figure}


\begin{references}

\bibitem{piezo_devices} K. Uchino, {\em Piezoelectric Actuators and
Ultrasonic Motors (Electronic Materials--Science \& Technology, 1),}
(Kluwer Academic Publishers, Boston, 1996).

\bibitem{ad} M.A. Akbas and P.K. Davies, 
{\em Solid State Chemistry of Inorganic Materials}, 
ed. P.K. Davies, A.J. Jacobson, C.C. Torardi, T and Vanderah, 
Proceedings {\bf 453}, 483 (1997).

\bibitem{cadp} L. Chai, M.A. Akbas, P.K. Davies, and J.B. Parise,
{\em Materials Research Bulletin} Pergamon {\bf 32} (1997).

\bibitem{ad2} M.A. Akbas and P.K. Davies, submitted as a Communication to 
{\em J. Am. Ceram. Soc.} (may, 1997).

\bibitem{ad3} M.A. Akbas and P.K. Davies,
{\em Journal of Materials Research.}  {\bf 12}, 2617 (1997).

\bibitem{Husson1} E. Husson,
{\em Mat. Res. Bull.}  {\bf 25}, 539 (1990).

\bibitem{Husson2} E. Husson,
{\em Mat. Res. Bull.}  {\bf 23}, 357 (1988).

\bibitem{Chen} J. Chen,
{\em J. Am. Cer. Soc.}  {\bf 72}, 593 (1989).

\bibitem{dcd} L. Dupont, L. Chai, P.K. Davis,
{\em Proceedings Materials Research Society, 
Solid State Chemistry of Inorganic Materials II,} {\bf 547}, 93-98 (1999).

\bibitem{ad4} M.A. Akbas and P.K. Davies,
{\em J. Am. Ceram. Soc.} {\bf 81}, 2205 (1998).

\bibitem{ad5} M.A. Akbas and P.K. Davies,
{\em J. Am. Ceram. Soc.} {\bf 81}, 1061 (1998).

\bibitem{mad} J.K. Montgomery, M.A. Akbas and P.K. Davies,
{\em J. Am. Ceram. Soc.} {\bf 82}, 3481 (1999).

\bibitem{ad2000} M.A. Akbas and P.K. Davies,
{\em J. Am. Ceram. Soc.} {\bf 83}, 119 (2000).

\bibitem{bv} L. Bellaiche and D. Vanderbilt,
{\em Phys. Rev. Lett.} {\bf 81}, 1318 (1998).

\bibitem{bc} B. Burton and E. Cockayne,
{\em Phys. Rev. B}  {\bf 60}, 12542 (1999).

\bibitem{wensell99} M. Wensell and H. Krakauer, AIP Proceedings.

\bibitem{wensell-2000} M. Wensell and H. Krakauer, to be published.

\bibitem{Schilfgaarde86} M. van Schilfgaarde, A.-B. Chen, and A. Sher,
{\em Phys. Rev. Lett.} {\bf 57}, 1149 (1986).

\bibitem{wei87} S.-H. Wei, 
{\em Phys. Rev. Lett.} {\bf 59}, 2613 (1987).

\bibitem{mwz} R. Magri, S.-H. Wei, A. Zunger,
{\em Phys. Rev. B}  {\bf 42}, 11388 (1990).

\bibitem{Metropolis} N. Metropolis {\it et al.},
{\em J. Chem. Phys.} {\bf 21}, 1087 (1953).

\bibitem{cohen_krakauer_92} R. E. Cohen and H. Krakauer, {\em
Ferroelectrics} {\bf 136}, 65 (1992).

\bibitem{xu2000} Z. Xu, S. M. Gupta, and D. Viehland, 
{\em J. Am. Ceram. Soc.} {\bf 83}, 181 (2000).

\end{references}
\end{document}